\documentclass[twocolumn,pra,aps,showpacs]{revtex4}

\usepackage{mathptmx}
\usepackage{subfigure}
\usepackage{psfrag,graphicx}
\usepackage{dcolumn}
\usepackage{amsmath,amssymb}
\usepackage{bm}
\usepackage{color}
\usepackage{latexsym}
\usepackage{epstopdf}
\usepackage{color}
\usepackage[english]{babel}
\usepackage{latexsym}
\usepackage{psfrag,graphicx}
\usepackage{subfigure}
\usepackage{amsmath}
\usepackage{amssymb}
\usepackage{amsfonts}
\usepackage{bm}
\usepackage{natbib}
\usepackage{epstopdf}
\usepackage{appendix}

\definecolor{mygrey}{gray}{0.35}
\definecolor{myblue}{rgb}{0.2,0.2,0.8}
\definecolor{myzard}{cmyk}{0,0,0.05,0}
\definecolor{mywhite}{rgb}{1,1,1}
\definecolor{mywhite}{rgb}{1,1,1}
\definecolor{myred}{rgb}{1,0.,0.3}

\usepackage[colorlinks=true,citecolor=myblue,linkcolor=myred]{hyperref}

\def\ba{\begin{align}}
\def\enda{\end{align}}
\def\bi{\begin{itemize}}
\def\ei{\end{itemize}}

\def\be{\begin{equation}}
\def\ee{\end{equation}}
\def\bea{\begin{eqnarray}}
\def\eea{\end{eqnarray}}
\def\bse{\begin{subequations}}
\def\ese{\end{subequations}}

\begin{document}
\title{Efficient approach for quantum sensing field gradients with trapped ions}
\author{Peter A. Ivanov}
\affiliation{Department of Physics and Research Center OPTIMAS, University of Kaiserslautern, Germany}
\affiliation{Department of Physics, St. Kliment Ohridski University of Sofia, James Bourchier 5 blvd, 1164 Sofia, Bulgaria}

\begin{abstract}
We introduce quantum sensing protocol for detection spatially varying fields by using two coupled harmonic oscillators as a
quantum probe. We discuss a physical implementation of the sensing technique with two trapped ions coupled via Coulomb mediated phonon hopping.
Our method relies on using the coupling between the localized ion oscillations and the internal states of the trapped ions which allows
to measure spatially varying electric and magnetic fields. First we discuss an adiabatic sensing technique which is capable to detect a
very small force difference simply by measuring the ion spin population. We also show that the adiabatic method can be used for detection magnetic
field gradient which is independent of the magnetic offset. Second, we show that the strong spin phonon coupling could leads to improve
sensitivity to force as well as to phase estimations. We quantify the sensitivity in terms of quantum Fisher information and show that it
diverges by approaching the critical coupling.

\end{abstract}

\maketitle


\section{Introduction}
The quantum sensors are of paramount importance in testing fundamental physics and quantum technologies. Examples include nanoscale mechanical oscillators for detecting very weak forces \cite{Teufel2009,Moser2013},
improved force microscope \cite{Butt2005} and magnetic field sensing using nuclei spins \cite{Jones2009}. In particular, trapped atomic ions provide an ideal platform for estimation of very weak signals due to the long coherence
time as well as the high accuracy control over the internal and vibrational states. On one hand the motion of the trapped ion can be approximated as harmonic mechanical oscillator with tunable mode frequency which allows to measure very weak electric fields \cite{Knunz2010}.
Recently, the detection of force that is off resonance with the trapping frequency of the singe trapped ion was experimentally demonstrated with force sensitivity in the range of aN ($10^{-18}$ N) per $\sqrt{\rm Hz}$
using Doppler velocimetry technique \cite{Shaniv2017}. Sensing of amplitude of motion in an ensemble of ions in Penning trap below the zero-point
fluctuations was demonstrated \cite{Gilmore2017}. On the other hand the internal states of the trapped ions can be used as a high sensitive
magnetometer for detecting very weak magnetic fields \cite{Kotler2014,Baumgart2016}.

Here we study the useful of the coupling between the individual motion modes of trapped ion system in the context of sensing spatially varying fields.
We consider a quantum probe system that consists of two harmonic mechanical oscillators coupled via Coulomb mediated phonon hopping.
Additionally we assume that external laser fields couple the localized ion oscillations which we refer to as local phonons with the internal states
of the trapped ions realizing the quantum Rabi lattice (RL) model. We propose sensing protocols capable to detect spatially varying electric
field that produce position depend displacement along the trap axis. First, we discuss adiabatic sensing protocol in which the spatially
varying field breaks the underlying parity symmetry of the RL model and as a result of that the force difference is detected
via measuring the spin population of one of the ions. We have shown earlier that the adiabatic protocol is very efficient technique for sensing
magnitude of the force with single trapped ion \cite{Ivanov2015,Ivanov2016}. Here we examine the ground-state order of the RL model and show
that nearest neighbor anti-ferromagnetic ground state allows to detect the linear combination of the magnitude of the forces along the trap axis.
Additionally, the information of the phase of the force can be extracted by controlling the laser phase and observing the coherent evolution of the
spin states. Furthermore, we show that the adiabatic sensing protocol can be applied for detecting magnetic field gradient which caused a
site-dependent atomic frequency shift. We show that thanks to the anti-ferromagnetic spin order the signal becomes independent on the offset
of the magnetic field.

Next, we discuss sensing protocol of the spatially varying electric field that relies on measuring the mean phonon number of the collective modes of
the two trapped ions. An estimation of the force magnitude via detecting the mean phonon number with single trapped ion was discussed
in \cite{IvanovPRA2016}. Here we show that the force difference caused excitation of phonons in the collective rocking mode while the mean phonon
number of the collective center-of-mass mode is proportional to the total force along the trap axis. We examine the force sensitivity as well as
the sensitivity in the estimation of the phase of the force in terms of quantum Fisher information. We show that the strong spin-boson coupling
leads to enhance sensitivity. Moreover, the corresponding quantum Fisher information diverges by approaching the critical coupling making the
quantum probe sensitive to infinitely small force perturbation. Finally, we discuss the optimal measurement that saturate the quantum Gramer-Rao bound.

The paper is organized as follows: For the sake of the reader's convenience, in Sec. \ref{model} we introduce the tight binding model which
described the Coulomb mediated coupling between the local phonons in the linear ion crystal. In Sec. \ref{RL_Imp} we discuss the physical
implementation of the quantum Rabi lattice model using linear ion crystal. In Sec. \ref{AP} we study the ground state order of the RL model
and discuss the adiabatic sensing protocol cabable to detect spatially varying electric or magnetic fields. In Sec. \ref{TIC} we extend the
sensing protocol to the three ion system. In Sec. \ref{CHO} we consider the limit of strong spin-boson coupling and discuss the sensitivity
in terms of quantum Fisher information. Finally, in Sec. \ref{conclusions} we summarize our findings.

\section{The Model}\label{model}
Our probe system consists of two ions with electric charge $e$ and mass $m$ confined in a linear Paul trap along the $z$ axis with trap
frequencies $\omega_{\beta}$ ($\beta=x,y,z$). The potential energy $\hat{V}$ of the system is a sum of the harmonic potential and the mutual
Coulomb repulsion between the ions in the trap \cite{Leibfried2003,Singer2010}
\begin{equation}
\hat{V}=\frac{m}{2}\sum_{\beta=x,y,z}\sum_{j=1}^{2}\omega_{\alpha}\hat{r}_{\alpha,j}^{2}+\frac{e^{2}}{|\hat{\vec{r}}_{1}-\hat{\vec{r}}_{2}|},
\end{equation}
where $\hat{\vec{r}}_{j}$ is the position vector operator of ion $j$. Assuming that the radial trap frequencies are much higher than the axial
trap frequency $(\omega_{x,y}\gg\omega_{z})$, the trapped ions are arranged in a linear configuration and occupy equilibrium position $z_{j}^{0}$
along the trap $z$ axis \cite{James1998}. Then, the position vector of ion $j$ can be expressed as $\hat{\vec{r}}=(z_{j}^{0}+\delta \hat{r}_{z,j})\vec{e}_{z}+\delta \hat{r}_{x,j}\vec{e}_{x}+\delta \hat{r}_{y,j}\vec{e}_{y}$,
where $\delta \hat{r}_{\alpha,j}$ are the displacement operators around the equilibrium positions. For a small displacement the vibrational degrees
of freedom in $x$, $y$ and $z$ direction are decoupled. In the following we only consider the oscillation in the transverse $x$ direction.
Treating the ions as an individual oscillators by introducing creation $\hat{a}_{j}^{\dag}$ and annihilation $\hat{a}_{j}$ operators of local
phonon at site $j$ such that $\hat{p}_{j}=i\sqrt{\hbar m\omega_{x}/2}(\hat{a}_{j}^{\dag}-\hat{a}_{j})$ and $\delta \hat{r}_{j}=\sqrt{\hbar/2m\omega_{x}}(\hat{a}_{j}^{\dag}+\hat{a}_{j})$, the vibrational Hamiltonian in the transverse $x$ direction becomes \cite{Porras2004}
\begin{eqnarray}
\hat{H}_{x}=\hbar\omega(\hat{a}_{1}^{\dag}\hat{a}_{1}+\hat{a}_{2}^{\dag}\hat{a}_{2})+\hbar \kappa(\hat{a}_{1}^{\dag}\hat{a}_{2}
+\hat{a}_{1}\hat{a}_{2}^{\dag}).\label{Hvib}
\end{eqnarray}
The Hamiltonian $\hat{H}_{x}$ is valid in the limit $\omega_{x}\gg \kappa$ which allows to neglect the phonon non-conserving
terms, $\hat{a}_{1}^{\dag}\hat{a}_{2}^{\dag}$ and $\hat{a}_{1}\hat{a}_{2}$. Within this approximation the local phonons are trapped with
renormalized phonon frequency $\omega$ and can hope between different sites with long-range hopping strength $\kappa$ where
\begin{equation}
\omega=\omega_{x}-\kappa,\quad \kappa=\frac{e^{2}}{2m\omega_{x}|\Delta z|^{3}},
\end{equation}
with $\Delta z=z_{2}^{0}-z_{1}^{0}$ being the distance between the two ions. Usually, the hopping amplitude is of the range of few kHz and
can be adjusted experimentally by controlling the axial trap frequency which vary the average ion distance. Recently, a radial phonon dynamics
subject to the tight-binding Hamiltonian (\ref{Hvib}) has been experimentally observed in a linear Paul trap \cite{Haze2012} as well as with
trapped ions in a double-well potential \cite{Harlander2011,Brown2011}. Moreover, the model (\ref{Hvib}) allows to study the strongly correlated
spin-boson phenomena such as quantum phase transition of polaritonic excitations \cite{Toyoda2013}, when external laser light is used to couple the
local phonon oscillations to the internal electronic states of the trapped ions. Recently, the experimental observation of single phonon
propagation in warm ion chain has been reported \cite{Abdelrahman2016}.

In the following we use the two coupled harmonic oscillators driven by external laser field as a highly sensitive probe for detecting spatially
varying electric and magnetic fields.

\section{Quantum Rabi Lattice Model as a quantum probe of gradient fields}\label{RL_Imp}
Our quantum probe system is represented by the quantum Rabi lattice model given by
\begin{eqnarray}
&&\hat{H}_{\rm RL}=\hat{H}_{x}+\hat{H}_{\rm s}+\hat{H}_{\rm sb},\quad \hat{H}_{\rm s}=\hbar\frac{\Omega}{2}\sum_{j=1}^{2}\sigma_{j}^{x},\notag\\
&&\hat{H}_{\rm sb}=\hbar \sum_{j=1}^{2}g_{j}(e^{i\phi_{j}}\hat{a}^{\dag}_{j}+e^{-i\phi_{j}}\hat{a}_{j})\sigma_{j}^{z},\label{RL}
\end{eqnarray}
where $\sigma_{j}^{\beta}$ are the Pauli operators for spin $j$. The term $\hat{H}_{\rm s}$ describes the interaction with external laser
field with Rabi frequency $\Omega$ which drives the transition between the spin states for each ion. The last term in (\ref{RL}) is the standard
spin-phonon dipolar coupling between the $j$th spin and the respective local phonon with coupling strength $g_{j}$ and phase $\phi_{j}$ which has been studied intensively in the context of quantum gate implementation and quantum simulation \cite{Schneider2012}.

Our goal is to measure spatially varying gradient fields, such as inhomogeneous electric field which produce different forces and respectively displacement along the ion chain or magnetic field gradient which causes spatially varying frequency splitting.

Let's consider first the physical implementation of the model (\ref{RL}). We assume that the effective spin states consist of the two metastable internal level of the ion $\left|\uparrow_{j}\right\rangle$, $\left|\downarrow_{j}\right\rangle$ separated by the frequency $\omega_{0}$. The interaction free Hamiltonian describing the system is $\hat{H}_{0}=\hat{H}_{x}+\hbar\omega_{0}\sum_{j=1}^{N}\sigma_{j}^{z}/2$. Consider that the ions simultaneously interact along the radial direction $x$ with two laser beams in a Raman configuration with laser frequency deference $\omega_{\rm L}=\omega-\delta$, where $\delta$ is a detuning with respect to the local phonon frequency $\omega$ ($\omega\gg\delta$). This interaction creates the coupling between the effective spin states and the local phonon oscillations with coupling strength $\Omega_{x,j}$.
Furthermore, additional field with frequency resonant with $\omega_{0}$ couples the spin states via carrier transition with Rabi frequency $\Omega$. The resulting interaction Hamiltonian reads
\begin{equation}
\hat{H}_{\rm_{I}}=\hbar\sum_{j=1}^{2}\{\Omega_{x,j}(e^{i\{ k_{x} \delta \hat{r}_{x,j}-\omega_{\rm L}t+\phi_{j}\}}+{\rm h.c.})\sigma_{j}^{z}+
\Omega\cos(\omega^{\prime}t)\sigma_{j}^{x}\},
\end{equation}
where $\vec{k}$ is the wave-vector difference pointing along the $x$ axis ($k_{x}=|\vec{k}|$), $\phi_{j}$ is the respective laser phase difference and $\omega^{\prime}$ is the frequency of the driving field which we assume to be in resonance with the atomic frequency $\omega^{\prime}=\omega_{0}$.
Next, we transform the total Hamiltonian $\hat{H}=\hat{H}_{0}+\hat{H}_{\rm I}$ into a rotating frame with respect to $\hat{U}_{R}(t)=e^{-i\omega_{0}t\sum_{j=1}^{2}\sigma_{j}^{z}-i(\omega-\delta)t\sum_{j=1}^{2}\hat{a}_{j}^{\dag}\hat{a}_{j}}$. Assuming the Lamb-Dicke limit and neglecting the fast oscillating terms we obtain
\begin{equation}
\hat{H}_{\rm_{RL}}=\hat{U}^{\dag}_{R}\hat{H}\hat{U}_{R}-i\hbar \hat{U}^{\dag}_{R}\partial_{t}\hat{U}_{R}
\label{Ht}
\end{equation}
where the spin-boson coupling is $g_{j}=\Omega_{x,j}\eta$ with $\eta=k_{x}x_{0}\ll 1$ being the Lamb-Dicke parameter and the local phonon frequency $\omega$ in Eq. (\ref{Hvib}) is replaced by the effective frequency $\delta$ in (\ref{Ht}). 
We note that recently a similar spin-boson interaction was used to measure experimentally a center-of-mass motion of two-dimensional
ion crystal in a Penning trap \cite{Gilmore2017}.
\subsection{Position-Dependent Kick}
In the first scenario consider here, we assume that the ions are exposed to a position dependent kick, which displace a motional amplitude according to $\hat{H}_{F}(t)=\sum_{j=1}^{N}\cos(\omega_{d}t+\xi)\vec{F}_{j}\cdot \hat{\vec{r}}_{j}$ where $\vec{F}_{j}$ is the force at the ion position $j$, $\omega_{d}$ is the oscillation frequency and $\xi$ is the phase. Consider that the frequency $\omega_{d}=\omega-\delta$ is closed to the local phonon frequency $\omega$, such that the only relevant displacement is along the $x$ axis. Transforming $\hat{H}_{F}(t)$ in to rotating frame with respect to $\hat{U}_{R}(t)$ and neglecting fast rotating terms we have
\begin{equation}
\hat{H}_{F}=\frac{F_{1}x_{0}}{2}(e^{i\xi}\hat{a}^{\dag}_{1}+e^{-i\xi}\hat{a}_{1})+\frac{F_{2}x_{0}}{2}(e^{i\xi}\hat{a}^{\dag}_{2}+e^{-i\xi}\hat{a}_{2}),\label{HF}
\end{equation}
where $F_{1}$ and $F_{2}$ are the parameters we wish to estimate. Moreover, we show that by controlling the laser phases we are able to detect
the phase of the force by observing the coherent evolution of the spin states or the mean phonon number. The total Hamiltonian becomes $\hat{H}=\hat{H}_{\rm RL}+\hat{H}_{F}$ where we treat $\hat{H}_{F}$ as a small symmetry-breaking perturbation.
\subsection{Magnetic-Field Gradient}
We can extend our sensing protocol for the case of static magnetic field gradient along the trapping axis $B=B_{0}+B^{\prime}z$, where $B_{0}$ is an offset field and $B^{\prime}$ is the constant gradient, the parameter we wish to estimate. Consider for example that our qubit states are formed by magnetic sensitive states. In the presence of magnetic field gradient the spin states of each ion exhibit a site-dependent resonance frequency \cite{Johanning2009}. The latter implies that the frequency $\omega^{\prime}$ which drives the carrier transition between the spin states will be slightly detune from the resonance such that $\omega^{\prime}=\omega_{0}-\delta B_{j}$, where $\delta B_{j}$ is the site-dependent detuning.  The resulting Hamiltonian in the rotating frame becomes $\hat{H}_{\rm s}\rightarrow\hat{H}_{\rm s}+\hat{H}_{B}$, where the symmetry-breaking term is
\begin{equation}
\hat{H}_{B}=\hbar(\delta B_{1}\sigma_{1}^{z}+\delta B_{2}\sigma_{2}^{z}).\label{HB}
\end{equation}
Here $\delta B_{j}=\lambda B_{0}+\lambda B^{\prime}z_{j}^{0}$ is the side-specific detuning with $\lambda$ being the coupling strength.

\section{Adiabatic force sensing protocol}\label{AP}
We begin by considering the adiabatic sensing protocol in which the transverse time-dependent Rabi frequency $\Omega(t)$ drives the system from normal phase to the symmetry-breaking phase where the effect of the perturbation term is estimated. We discuss the energy spectrum of Hamiltonian (\ref{RL}) in the limit $\Omega=0$ such that $\hat{H}_{\rm s}=0$. Moreover, we treat $\hat{H}_{F}$ as a perturbation term which is valid as long as $\omega_{c(r)},J\gg F_{j}x_{0}/2\hbar$. In that case the model can be treated exactly. Indeed, let us introduce the operators $\hat{a}_{ c}=(\hat{a}_{1}+\hat{a}_{2})/\sqrt{2}$ and $\hat{a}_{r}=(\hat{a}_{1}-\hat{a}_{2})/\sqrt{2}$ that annihilate collective phonon excitation in the center-of-mass mode and respectively in the rocking mode. Assuming equal couplings $g_{j}=g$ and laser phases $\phi_{j}=\phi$ the Hamiltonian takes the form
\begin{eqnarray}
\hat{H}_{\rm RL}&=&\hbar\omega_{c}\hat{a}_{c}^{\dag}\hat{a}_{c}+\hbar\omega_{r}\hat{a}_{r}^{\dag}\hat{a}_{r}+
\frac{\hbar g}{\sqrt{2}}\{\sigma_{1}^{z}(\hat{a}_{c}^{\dag}+\hat{a}_{r}^{\dag}+\hat{a}_{c}+\hat{a}_{r})\notag\\
&&+\sigma_{2}^{z}(\hat{a}_{c}^{\dag}-\hat{a}_{r}^{\dag}+\hat{a}_{c}-\hat{a}_{r})\},
\end{eqnarray}
where
\begin{equation}
\omega_{c}=\delta+\kappa,\quad \omega_{r}=\delta-\kappa,\label{frequencies}
\end{equation}
are the center-of-mass and rocking mode frequencies. Next, we perform canonical transformation such that $\hat{\tilde{H}}=\hat{D}^{\dag}(\hat{\alpha}_{c})\hat{D}^{\dag}(\hat{\alpha}_{r})\hat{H}_{\rm RL}\hat{D}(\hat{\alpha}_{r})\hat{D}(\hat{\alpha}_{c})$, where $\hat{D}(\hat{\alpha}_{q})=e^{\hat{\alpha}_{q}\hat{a}^{\dag}_{q}-\hat{\alpha}^{\dag}_{q}\hat{a}_{q}}$ $q=c,r$ is the displacement operator with $\hat{\alpha}_{c}=-\frac{g}{\sqrt{2}\omega_{c}}(\sigma_{1}^{z}+\sigma_{2}^{z})$ and $\hat{\alpha}_{r}=\frac{g}{\sqrt{2}\omega_{r}}(\sigma_{2}^{z}-\sigma_{1}^{z})$. The transformed Hamiltonian becomes
\begin{equation}
\hat{\tilde{H}}=\hbar\omega_{c}\hat{a}_{c}^{\dag}\hat{a}_{c}+\hbar\omega_{r}\hat{a}_{r}^{\dag}\hat{a}_{r}+
\hbar J\sigma_{1}^{z}\sigma_{2}^{z},\label{Htran}
\end{equation}
where $J=g^{2}\left(\frac{1}{\omega_{r}}-\frac{1}{\omega_{c}}\right)$ is the spin-spin coupling and we have omitted the constant term. The energy spectrum of Hamiltonian (\ref{Htran}) is a double-degenerate with eigenvectors $|\Phi_{s_{1},s_{2},n_{c},n_{b}}\rangle=|s_{1},s_{2}\rangle|n_{c}\rangle|n_{r}\rangle$ ($s=\uparrow,\downarrow$) and energies $E_{\pm,n_{c},n_{r}}=\hbar n_{c}+\hbar n_{r}\pm\hbar J$. Here the state $|n_{c}\rangle|n_{r}\rangle$ describes a Fock state with $n_{c}$ phonons in the the center-of-mass mode and respectively $n_{r}$ phonon in the rocking mode.
Because in the radial direction the center-of-mass mode is the highest energy mode ($\omega_{c}>\omega_{r}$) \cite{Marquet2003} we have $J>0$. The latter implies that the double degenerate ground states of $\hat{H}_{\rm RL}$ in the limit $\Omega=0$ correspond to an anti-ferromagnetic spin order, namely
\begin{equation}
|\Psi_{+}\rangle=\left|\downarrow\uparrow\right\rangle|0_{c}\rangle|\alpha_{r}\rangle,\quad |\Psi_{-}\rangle=\left|\uparrow\downarrow\right\rangle|0_{c}\rangle|-\alpha_{r}\rangle,\label{ground}
\end{equation}
with vibrational ground state in the center-of-mass mode and displaced motion state with amplitude $\alpha_{r}=\sqrt{2}g/\omega_{r}$ in the rocking mode. The other set of infinitely many double-degenerate excited states consists of the states $|\Phi_{-,n_{c},n_{r}}\rangle=\left|\downarrow\downarrow\right\rangle \hat{D}(\alpha_{c})|n_{c}\rangle|n_{r}\rangle$ and $|\Phi_{+,n_{c},n_{r}}\rangle=\left|\uparrow\uparrow\right\rangle \hat{D}(-\alpha_{c})|n_{c}\rangle|n_{r}\rangle$ with displacement amplitude $\alpha_{c}=\sqrt{2}g/\omega_{c}$.

Next, we treat $\hat{H}_{\rm s}$ as a perturbation term, which lifts the degeneracy of the ground-state manifold by creating equal entangled superposition between the states (\ref{ground}). Since, $\langle \Psi_{+}|\hat{H}_{\rm_{s}}|\Psi_{-}\rangle=0$, the effective coupling is induced by the second order processes, which gives $\hat{H}_{\rm eff}=-\Delta_{\rm c} \tilde{\sigma}_{x}$, where $\tilde{\sigma}_{x}=|\Psi_{+}\rangle\langle\Psi_{-}|+|\Psi_{-}\rangle\langle\Psi_{+}|$. Assuming $\delta>g$ the collective modes are only virtually excited which allows to obtain $\Delta_{\rm c}=(\Omega^{2}/4J)e^{-|\alpha_{c}|^{2}-|\alpha_{r}|^{2}}$. Hereafter we assume that the transverse field vary in time as $\Omega(t)=\Omega(0)e^{-\gamma t}$ which implies $\Delta_{\rm c}\sim e^{-2\gamma t}$.
\begin{figure}
\includegraphics[width=0.45\textwidth]{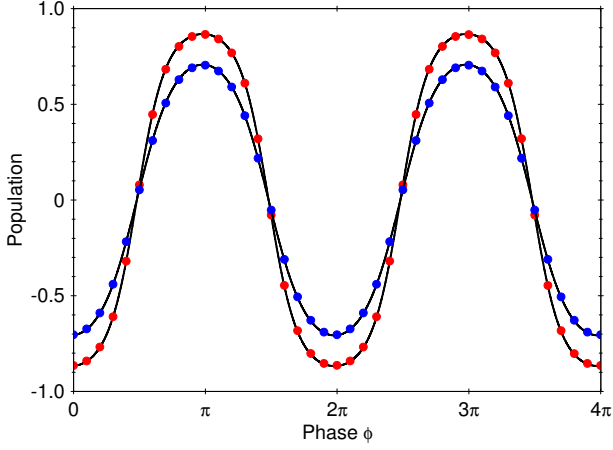}
\caption{(Color online) The expectation values of $\langle\sigma_{1}^{z}(t_{f})\rangle=p_{\uparrow}(t_{f})-p_{\downarrow}(t_{f})$ versus the externally controlled laser phase $\phi$. We assume $\Omega=825$ kHz, $\gamma=0.1$ kHz, $\delta=70$ kHz, $\kappa=12$ kHz, $\xi=0.98\pi$ and $x_{0}=14.5$ nm. The solid lines represent the analytical expression (\ref{signal}) while the red dots ($F_{1}=3.78$ yN, $F_{2}=0.95$ yN) and blue triangle ($F_{1}=2.84$ yN, $F_{2}=0.95$ yN) are the numerical solution of the time-dependent Schr\"odinger equation with Hamiltonian $\hat{H}_{\rm RL}(t)+\hat{H}_{F}$. }
\label{fig1}
\end{figure}

The sensing protocol starts by preparing the system initially in the state $|\psi(0)\rangle=\left|--\right\rangle|0_{c}\rangle|0_{r}\rangle$ ($\sigma_{x}|\pm\rangle=\pm|\pm\rangle$) which is the the ground state of Hamiltonian (\ref{RL}) in the limit $\Omega(0)\gg \omega_{c},g$. Then, the field $\Omega(t)$ decreases with the time such that the system evolves into the entangled state $|\psi(t)\rangle=c_{+}(t)|\Psi_{+}\rangle+c_{-}(t)|\Psi_{-}\rangle$, where $c_{\pm}(t)$ are the respective probability amplitudes. The effect of symmetry-breaking perturbations terms $\hat{H}_{F}$ and $\hat{H}_{B}$ is to create non-equal superposition between the ground-state manifold states with $|c_{+}(t_{f})|^{2}\neq|c_{-}(t_{f})|^{2}$. By solving the two-state problem with effective Hamiltonian $\hat{H}_{\rm eff}=-\Delta_{\rm c}(t)\tilde{\sigma}_{x}+A_{F(B)}\tilde{\sigma}_{z}$, one could derive analytical expressions for the probability amplitudes $c_{\pm}(t_{f})$, see Appendix \ref{STSP}. Here $A_{F(B)}=\langle \Psi_{+}|\hat{H}_{F(B)}|\Psi_{+}\rangle-\langle \Psi_{-}|\hat{H}_{F(B)}|\Psi_{-}\rangle$ are the matrix elements of the respective perturbation terms (\ref{HF}) and (\ref{HB}) within the ground state manifold. For the specific time-dependence of $\Omega(t)$ the Hamiltonian $\hat{H}_{\rm eff}$ is reduced to the Demkov model which is exactly solvable \cite{Ivanov2015,Vitanov}.
\subsection{Sensing Position Dependent Kick}
For the case of force symmetry-breaking term (\ref{HF}) we obtain $\langle \Psi_{\pm}|\hat{H}_{F}|\Psi_{\pm}\rangle=\pm\cos(\phi-\xi)\frac{gx_{0}}{\omega_{r}}(F_{1}-F_{2})$ which implies that the measured signal at time $t_{f}\gg\gamma^{-1}$ depends only on the force difference
\begin{eqnarray}
&&p_{\uparrow}(t_{f})=\frac{1}{2}+\frac{1}{2}\tanh\left(\frac{\pi gx_{0}\cos(\phi-\xi)F_{-}}{2\hbar\gamma\omega_{r}}\right),\notag\\
&&\langle\sigma_{1}^{z}(t_{f})\rangle=2p_{\uparrow}(t_{f})-1,\label{signal}
\end{eqnarray}
where $F_{-}=F_{1}-F_{2}$. Due to the anti-ferromagnetic spin order the expectation value of the spin states for the second ion is $\langle\sigma_{2}^{z}(t_{f})\rangle=-\langle\sigma_{1}^{z}(t_{f})\rangle$. The result (\ref{signal}) shows that by measuring the spin population of one of the ions via state dependent fluorescence technique one could determine the parameter $\xi$ by varying the externally controlled laser phase $\phi$, until the oscillation amplitude vanishes. Moreover, Eq. (\ref{signal}) allows also to determine the force difference $F_{-}$ from the same signal versus $\phi$, which is related with the oscillation amplitude as is shown in Fig. \ref{fig1}. We point out that the uncertainty of the joint estimation of the both parameters is however unbounded since the two parameter estimation requires measurement with at least three outputs \cite{Vidrighin2014}. Thus the detection of the force difference requires a prior knowledge of the phase $\xi$ and vice versa.
Alternatively, we could address globally the ion chain and measure the probability of the collective states. Within the anti-ferromagnetic ground state manifold, the respective probabilities become $p_{\downarrow\downarrow}(t_{f})=p_{\uparrow\uparrow}(t_{f})=0$, $p_{\uparrow\downarrow}(t_{f})=\frac{1}{2}(1+\langle\sigma_{1}^{z}(t_{f})\rangle)$ and $p_{\downarrow\uparrow}(t_{f})=\frac{1}{2}(1-\langle\sigma_{1}^{z}(t_{f})\rangle)$.

The error in the estimation of the parameters $F_{-}$ and $\xi$ is bounded by the Cramer-Rao inequality \cite{Toth2014}
\begin{equation}
\Delta^{2}\varphi\geq\frac{1}{N_{\rm exp}I_{\rm cl}(\varphi)},
\end{equation}
where $\varphi$ is either $F_{-}$ or $\xi$, $N_{\rm exp}$ is the number of experimental repetitions and $I_{\rm cl}(\varphi)=\sum_{n}\frac{1}{p_{n}}\left(\frac{\partial p_{n}}{\partial\varphi}\right)^{2}$ is the classical Fisher information. Using Eq. (\ref{signal}) we find
\begin{equation}
I_{\rm cl}(F_{-})=\left(\frac{\pi g x_{0}}{2\hbar\gamma\omega_{r}}\right)^{2}\frac{\cos^{2}(\phi-\xi)}{\cosh^{2}\left(\frac{\pi gx_{0}\cos(\phi-\xi)F_{-}}{2\hbar\gamma\omega_{r}}\right)},\label{Fisher_F}
\end{equation}
and respectively
\begin{equation}
I_{\rm cl}(\xi)=\left(\frac{\pi g x_{0}}{2\hbar\gamma\omega_{r}}\right)^{2}\frac{F^{2}_{-}\sin^{2}(\phi-\xi)}{\cosh^{2}\left(\frac{\pi gx_{0}\cos(\phi-\xi)F_{-}}{2\hbar\gamma\omega_{r}}\right)}.\label{Fisher_Phase}
\end{equation}
The result (\ref{Fisher_F}) shows that the best sensitivity for force difference detection is achieved for $\phi=\xi$ where the signal has a maximum, while the phase estimation error is minimal for $\phi\approx \xi+\frac{\pi}{2}$ where the slope of the signal is sharpest, see Fig. \ref{fig1}.

Alternatively, as a figure of merit for the sensitivity we can use the signal-to-noise ratio ${\rm SNR}=\langle\sigma_{j}^{z}(t_{f})\rangle/\langle\Delta^{2}\sigma_{j}^{z}(t_{f})\rangle^{1/2}$ which is larger for better estimation. For $\phi=\xi$, the minimal detectable difference corresponding to a signal-to-noise ratio of one is
\begin{equation}
F_{-}^{\rm min}=\frac{2\hbar \gamma(\delta-\kappa)}{\pi g x_{0}}\sinh^{-1}(1),
\end{equation}
where $\langle\Delta^{2}\sigma_{j}^{z}(t_{f})\rangle=1-\langle\sigma_{j}^{z}(t_{f})\rangle^{2}$ is the variance of the signal. For example, using the parameters in Fig. \ref{fig1} the minimal detectable force difference correspond approximately to $F_{-}^{\rm min}\approx1.9$ yN ($10^{-24}$ N).

Finally we point out that the measurement strategy based on the local detection of the spin population of one of the two ions is optimal. Indeed, the ground-state anti-ferromagnetic spin order implies that $\langle\sigma_{j}^{x}\rangle=\langle\sigma_{j}^{y}\rangle=0$ such that only measurements in the basis of the $\sigma_{z}$ operator give non-zero signal.

On the other hand the $\rm SNR$ is bounded by the quantum Fisher information $I_{Q}(\varphi)$ which doest not depend on the specific measurement being performed and gives the ultimate bounds of the estimation precision \cite{Paris2009}
\begin{equation}
\Delta^{2}\varphi\geq\frac{1}{N_{\rm exp}I_{Q}(\varphi)}\label{QCR}
\end{equation}
and $I_{\rm cl}(\varphi)\leq I_{Q}(\varphi)$. For pure state it reads
\begin{equation}
I_{Q}(\varphi)=4\{\langle \partial_{\varphi}\psi|\partial_{\varphi}\psi\rangle-|\langle \psi|\partial_{\varphi}\psi\rangle|^{2}\}.
\end{equation}
Using the solution of the two-state problem one can derive analytical expression for the quantum Fisher information (see, Appendix \ref{STSP})
\begin{equation}
I_{Q}(F_{-})=\left(\frac{g x_{0}}{2\hbar\gamma\omega_{r}}\right)^{2}\frac{\pi^{2}+4(\ln(z)-\Re \Psi(\beta))^{2}}{\cosh^{2}\left(\frac{\pi gx_{0}F_{-}}{2\hbar\gamma\omega_{r}}\right)},\label{QFIFD}
\end{equation}
where $z=(x/2) e^{-2\gamma t_{f}}$ with $x=\Omega^{2}/8J\gamma$ and $\Psi(\beta)$ is the digamma function with $\beta=\frac{1}{2}+i\frac{ gx_{0}F_{-}}{2\hbar\gamma\omega_{r}}$, which shows that the estimation precision increases in time as $t^{2}$. Similar expression can be derived for the phase estimation, see Appendix \ref{STSP}.
\subsection{Sensing Magnetic-Field Gradient}
\begin{figure}
\includegraphics[width=0.45\textwidth]{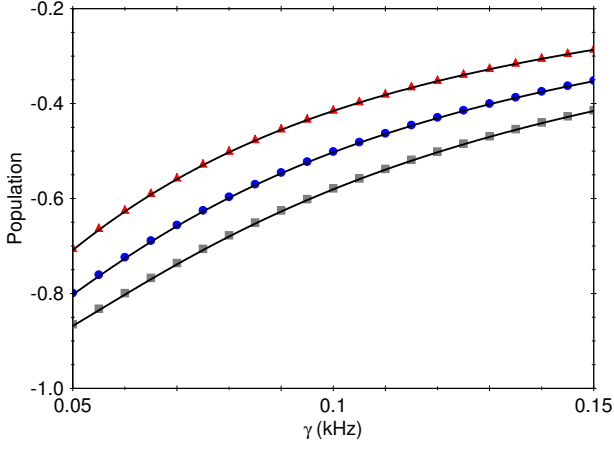}
\caption{(Color online) The expectation value of $\langle\sigma_{1}^{z}(t_{f})\rangle$ versus the slope $\gamma$. We assume $\Omega=925$ kHz, $g=25$ kHz, $\delta=50$ kHz, $\kappa=11$ kHz, $\Delta z=4$ $\mu$m and $g_{J}=2$. The solid lines are the analytical expression (\ref{signalB}) and the red triangles $B^{\prime}=4\times10^{-11}$ T/$\mu$m, blue dots  $B^{\prime}=5\times10^{-11}$ T/$\mu$m and gray squares $B^{\prime}=6\times10^{-11}$ T/$\mu$m are the numerical solution of the time-dependent Schr\"odinger equation with Hamiltonian $\hat{H}_{\rm RL}(t)+\hat{H}_{B}$. }
\label{figmag}
\end{figure}
The same technique can be applied for sensing magnetic field gradient. In that case the matrix elements of the symmetry-breaking term $\hat{H}_{B}$ within the ground-state manifold are $\langle \Psi_{\pm}|\hat{H}_{B}|\Psi_{\pm}\rangle=\pm\hbar\lambda B^{\prime}\Delta z$. Consequently, the signal at time $t_{f}$ is independent on the offset field $B_{0}$ and depends only on the magnitude of the magnetic field gradient
\begin{equation}
\langle\sigma_{1}^{z}(t_{f})\rangle=-\tanh\left(\frac{\pi \lambda B^{\prime}\Delta z}{2\gamma}\right),\label{signalBB}
\end{equation}
and $\langle\sigma_{2}^{z}(t_{f})\rangle=-\langle\sigma_{1}^{z}(t_{f})\rangle$. In Fig. \ref{figmag} we compare the exact result of the signal at $t_{f}$ versus the slope $\gamma$ with analytical expression (\ref{signalBB}), where perfect agreement is observed. The minimal detectable magnetic field gradient corresponding to signal-to-noise ration of one is
\begin{equation}
B^{\prime}_{\rm min}=\frac{2\gamma}{\pi\lambda\Delta z}\sinh^{-1}(1).
\end{equation}
The coupling strength is $\lambda=g_{J}\mu_{\rm B}/\hbar$, where $g_{J}$ is is the Lande $g$ factor and $\mu_{\rm B}$ is the Bohr magneton. Assuming a distance between the two ions $\Delta z=4$ $\mu$m and $\gamma=0.05$ kHz we estimate minimal detectable magnetic-field gradient $B_{\rm min}^{\prime}\approx 4\times 10^{-11}$ T/$\mu$m.
\subsection{Ferromagnetic Spin Order}
By controlling the individual spin phonon couplings $g_{j}$ one could obtain a ferro-magnetic spin order as a ground state of Hamiltonian (\ref{RL}) for $\Omega=0$. Indeed, if we set $g_{1}=-g_{2}=g$, then the corresponding Hamiltonian after making displacement transformation is identical to (\ref{Htran}) by replacing the spin-spin coupling $J$ by $-J$. Consequently, the the ground-state manifold supports ferromagnetic spin order,
\begin{equation}
|\Phi_{+}\rangle=\left|\downarrow\right\downarrow\rangle|0_{c}\rangle|\alpha_{r}\rangle,\quad |\Phi_{-}\rangle=\left|\uparrow\uparrow\right\rangle|0_{c}\rangle|-\alpha_{r}\rangle,\label{ferro}
\end{equation}
such that the system evolves into a superposition state $|\psi(t)\rangle=c_{+}(t)|\Phi_{+}\rangle+c_{-}(t)|\Phi_{-}\rangle$. The force measurement is not affected by the ferro-magnetic spin order because the matrix elements of the force term $\hat{H}_{F}$ in the basis (\ref{ferro}) are the same as the matrix elements within the anti-ferromagnetic manifold. As a result of that the measured signal is given by Eq. (\ref{signal}). However, for the perturbation $\hat{H}_{B}$ (\ref{HB}) we obtain $\langle \Phi_{\pm}|\hat{H}_{B}|\Phi_{\pm}\rangle=\mp\hbar(\delta B_{1}+\delta B_{2})$ such that the signal at $t_{f}$ is
\begin{equation}
\langle\sigma_{1}^{z}(t_{f})\rangle=-\tanh\left(\frac{\pi (\delta B_{1}+\delta B_{2})}{2\gamma}\right),\label{signalB}
\end{equation}
which depends on the sum of the magnetic field gradient at the two ion's positions.
\section{Three Ion Case}\label{TIC}
\begin{figure}
\includegraphics[width=0.45\textwidth]{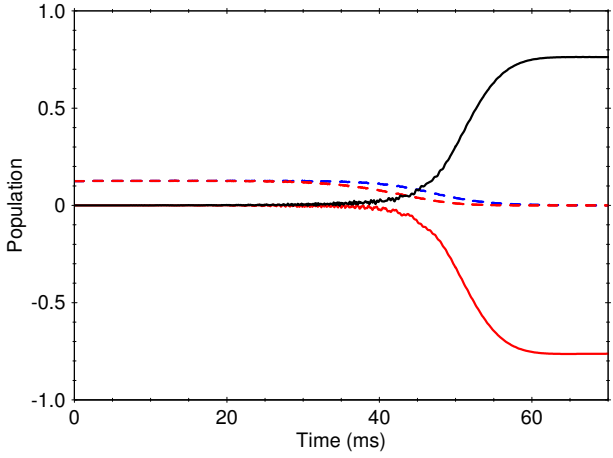}
\caption{(Color online) The exact numerical result for the expectation value $\langle\sigma_{1}^{z}(t)\rangle$ (black curve) and $\langle\sigma_{2}^{z}(t)\rangle$ (red curve) as a function of time for ion chain with three ions. According the ground-state order (\ref{n3spin}) we have $\langle\sigma_{3}^{z}(t)\rangle=\langle\sigma_{1}^{z}(t)\rangle$. The red dashed line $p_{\downarrow\downarrow\downarrow}(t)$ and blue dashed line $p_{\downarrow\downarrow\uparrow}(t)$ are the collective probabilities which tend to zero as the time increases, indicating the nearest neighbour anti-ferromagnetic spin state. The signal asymptotically tend to Eg. (\ref{signaln3}). The parameters are set to $F_{1}=4.3$ yN, $F_{2}=1.3$ yN, $F_{3}=1.0$ yN, $\delta=45$ kHz, $g=5$ kHz, $\kappa=12$ kHz, $\Omega=2.73$ MHz and $\gamma=0.13$ kHz. We assume non-zero phases  $\xi=1.2\pi$ and $\phi=0.9\pi$.   }
\label{fign3}
\end{figure}
The adiabatic sensing protocol is not restricted to two ion case but can be extended for higher number of ions which allows to measure a linear combination of force magnitudes or magnetic-field gradient along the ion chain. Here we consider the case of three trapped ions with nearest-neighbour hopping.
Setting $g_{1}=g_{3}=g$ and $g_{2}=\sqrt{2}g$ the Rabi lattice Hamiltonian can be brought after canonical transformation into the form (see Appendix \ref{TIC})
\begin{eqnarray}
\hat{\tilde{H}}&=&\hbar\omega_{c}\hat{a}_{c}\hat{a}_{c}+\hbar\omega_{r}\hat{a}_{r}\hat{a}_{r}+\hbar\omega_{e}\hat{a}^{\dag}_{e}\hat{a}_{e}
+\hbar J(\sigma_{1}^{z}\sigma_{2}^{z}+\sigma_{2}^{z}\sigma_{3}^{z})\notag\\
&&+\hbar J^{\prime}\sigma^{z}_{1}\sigma^{z}_{3},
\end{eqnarray}where we have introduced collective modes with frequencies $\omega_{c}=\delta+\sqrt{2}\kappa$, $\omega_{r}=\delta-\sqrt{2}\kappa$
and $\omega_{e}=\delta$. We find that the nearest-neighbour spin-spin coupling is positive $J=g^{2}(\frac{1}{\omega_{r}}-\frac{1}{\omega_{c}})$,
while the next neighbour spin-spin coupling is negative $J^{\prime}=g^{2}(\frac{1}{\omega_{e}}-\frac{1}{2\omega_{r}}-\frac{1}{2\omega_{c}})$.
As a consequence of that the double-degenerate ground-state spin configuration that minimize all spin-spin interactions is
\begin{equation}
|\Psi_{+}\rangle=\left|\downarrow\uparrow\downarrow\right\rangle|0_{c}\rangle|\alpha_{r}\rangle|0_{e}\rangle,\quad
|\Psi_{-}\rangle=\left|\uparrow\downarrow\uparrow\right\rangle|0_{c}\rangle|-\alpha_{r}\rangle|0_{e}\rangle,\label{n3spin}
\end{equation}
where $|\alpha_{r}\rangle$ is a coherent state with $\alpha_{r}=2g/\omega_{r}$. The adiabatic sensing protocol starts by preparing the system into the initial state $|\psi(0)\rangle=\left|---\right\rangle|0_{c}\rangle|0_{r}\rangle|0_{e}\rangle$ for $\Omega(0)\gg\omega_{c},g$ which evolves into the entangled state $|\psi(t_{f})\rangle=c_{+}|\Psi_{+}\rangle+c_{-}(t_{f})|\Psi_{-}\rangle$. Measuring the spin population of the first ion we find
\begin{eqnarray}
&&p_{\uparrow}(t_{f})=\frac{1}{2}+\frac{1}{2}\tanh\left(\frac{\pi gx_{0}\cos(\phi-\xi)F^{\prime}_{-}}{3\hbar\gamma\omega_{r}}\right)\notag\\
&&F^{\prime}_{-}=F_{1}-\sqrt{2}F_{2}+F_{3}.\label{signaln3}
\end{eqnarray}
and respectively $\langle\sigma_{1}^{z}(t_{f})\rangle=-\langle\sigma_{2}^{z}(t_{f})\rangle$ as is shown in Fig. (\ref{fign3}).
\begin{figure}
\includegraphics[width=0.45\textwidth]{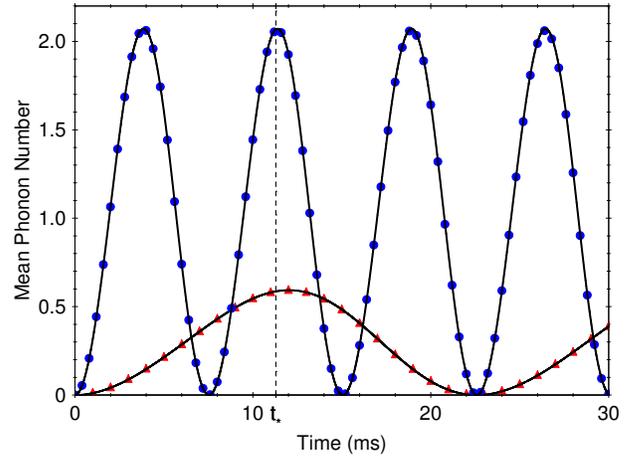}
\caption{(Color online) The mean phonon number as a function of time. The parameter as set to $F_{1}=7$ yN, $F_{2}=5$ yN, $\delta=0.6$ kHz, $g=2.5$ kHz, $\Omega=300$ kHz, $\phi=\pi/3$, and $\xi=\pi/2$. The solid lines are the analytical expression for the signal (\ref{signal_n}). The red triangles (rocking mode) and blue dots (center-of-mass mode) are the numerical solution of the time-dependent Schr\"odinger equation with Hamiltonian $\hat{H}_{\rm RL}+\hat{H}_{F}$. The vertical dashed line depicts the time $t_{*}$ at which the both mean phonon number are equal to $\langle \hat{a}_{q}^{\dag}\hat{a}_{q}\rangle=4|\alpha_{q}|^{2}$ which is fulfilled for hopping amplitude $\kappa_{*}=0.28$ kHz.}
\label{fig2}
\end{figure}
\section{Coupled Harmonic Oscillators}\label{CHO}
In the following we discuss measurement protocol for spatially varying forces with oscillation frequency close to resonance with respect to the frequency of the two coupled harmonic oscillators with small detuning $\delta\ll g$.
We show that the strong spin-boson interaction is an essential for the sensing protocol in way that the quantum Fisher information diverges at the critical spin-boson coupling making the quantum probe sensitive to very small symmetry-breaking force perturbation.
Such a criticality of the quantum Fisher information in the estimation of the parameter that drives the dynamics of the many-body systems was study in \cite{Zanardi2008}. It was shown that the estimation precision is enhanced by approaching the critical coupling in many-body systems exhibiting quantum phase transitions. Examples include high-precision estimation of the coupling in the Dicke model closed to the normal-to-superradiant phase transition \cite{Bina2016} as well as finite-temperature estimation of the anisotropy parameter in the Lipkin-Meshkov-Glick critical system \cite{Salvatori2014}. Here, our system is finite and the criticality of the system is
controlled by parameter that does not depend on the perturbation term and can be tuned for example by adjusting the effective phonon frequency. In contrast with the previous adiabatic scheme, here we assume that the transverse field $\Omega$ is a constant in time and $\Omega\gg \delta,g$ such that the spin-degree of freedom in the Rabi lattice model (\ref{RL}) can be traced out which leads to pure coupled bosonic model. The latter implies that the sensing protocol is not capable to detect magnetic field gradient because the spin degree of freedom are effectively frozen. Thus, hereafter we focus on sensing protocol for detecting spatially varying displacement via measuring the mean phonon numbers in the collective vibrational modes \cite{Maiwald2009}.
\begin{figure}
\includegraphics[width=0.45\textwidth]{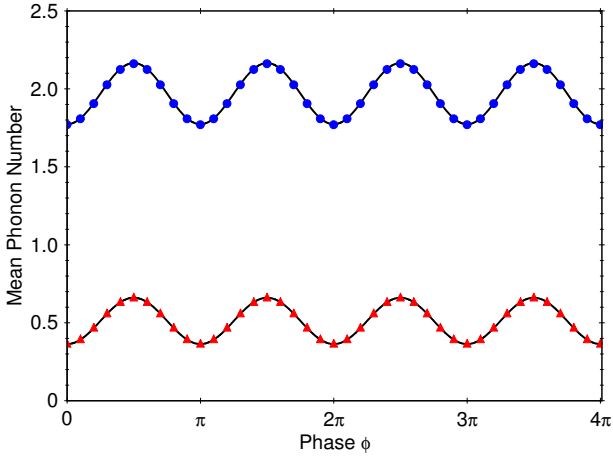}
\caption{(Color online) The mean phonon number versus the laser phase $\phi$ at time $t_{*}$ where the signal becomes $\langle \hat{a}_{q}^{\dag}\hat{a}_{q}\rangle=4|\alpha_{q}|^{2}$ with displacement parameter $|\alpha_{q}|$ given by Eq. (\ref{displacement}). The solid lines are the analytical results, while the red triangles (rocking mode) and blue dots (center-of-mass) mode are the exact results. The signals review oscillatory behaviour with amplitude proportional to the force difference $F_{-}=F_{1}-F_{2}$ for the rocking mode and respectively to the total force $F_{c}=F_{1}+F_{2}$ for the center-of-mass mode. The parameters are the same as in Fig. \ref{fig2} }
\label{fig3}
\end{figure}

Let's perform canonical transformation $\hat{U}=e^{\hat{S}}$ with
\begin{equation}
\hat{S}=i\frac{g}{\Omega}\sum_{j=1}^{2}\sigma_{j}^{y}(\hat{a}^{\dag}_{j}e^{i\phi}+\hat{a}_{j}e^{-i\phi}),
\end{equation}
which leads to an effective Hamiltonian $\hat{H}_{\rm eff}=e^{-\hat{S}}(\hat{H}_{\rm RL}+\hat{H}_{F})e^{\hat{S}}$, namely
\begin{eqnarray}
&&\hat{H}_{\rm eff}=\hat{H}_{x}+\hat{H}_{\rm s}+\hat{H}_{\rm b}+\hat{H}_{F},\notag\\
&&\hat{H}_{\rm b}=\hbar\frac{g^{2}}{\Omega}\sum_{j=1}^{2}\sigma_{j}^{x}(\hat{a}^{\dag}_{j}e^{i\phi}+\hat{a}_{j}e^{-i\phi})^{2},\label{Hboson}
\end{eqnarray}
where we keep only terms of order of $g^{2}/\Omega$. We observe that the Hamiltonian (\ref{Hboson}) is diagonal in the spin basis $\{|-\rangle,|+\rangle\}$ and thus the Hilbert space is decomposed into four orthogonal subspaces corresponding to each of the spin configurations. In the following we assuming that the both spins are initially prepared in the state $\left|\psi_{\rm s}(0)\right\rangle=\left|--\right\rangle$ such that the effective Hamiltonian becomes
\begin{eqnarray}
\hat{H}_{\rm eff}&=&\hbar\tilde{\delta}(\hat{a}^{\dag}_{1}\hat{a}_{1}+\hat{a}^{\dag}_{2}\hat{a}_{2})
+\hbar\kappa(\hat{a}_{1}^{\dag}\hat{a}_{2}+\hat{a}_{1}\hat{a}_{2}^{\dag})\notag\\
&&-\frac{\hbar g^{2}}{\Omega}\sum_{j=1}^{2}(e^{2i\phi}\hat{a}^{\dag2}_{j}+{\rm h.c.})
+\sum_{j=1}^{2}\frac{F_{j}x_{0}}{2}(e^{i\xi}\hat{a}^{\dag}_{j}+{\rm h.c.}),\label{h}
\end{eqnarray}
where $\tilde{\delta}=\delta(1-\zeta^{2}/2)$ with $\zeta^{2}=4g^{2}/\Omega\delta$ and we have omitted the constant term. Next, we introduce collective center-of-mass and rocking mode operators $\hat{a}_{q}$ ($q=c,r$), which decouple Hamiltonian (\ref{h}) into two uncoupled oscillators with new renormalized frequencies $\tilde{\omega}_{q}=\omega_{q}(1-\zeta_{q}^{2}/2)$ where $\zeta_{q}^{2}=4g^{2}/\Omega\omega_{q}$ and collective frequencies $\omega_{q}$ given by Eq. (\ref{frequencies}). We find that the unitary propagator corresponding to Hamiltonian (\ref{h}) can be expressed as $\hat{U}(t)=\hat{U}_{c}(t)\hat{U}_{r}(t)$ with
\begin{equation}
\hat{U}_{q}=\hat{D}^{\dag}(\alpha_{q})\hat{S}^{\dag}(\nu_{q})e^{-i\vartheta_{q}t\hat{a}^{\dag}_{q}\hat{a}_{q}}
\hat{S}(\nu_{q})\hat{D}(\alpha_{q}).
\end{equation}
where $\vartheta_{q}=\omega_{q}\sqrt{1-\zeta_{q}^{2}}$ is the force-independent phase, $\hat{S}(\nu_{q})=e^{\nu_{q}(\hat{a}^{\dag 2}_{q}-\hat{a}_{q})}$ is the squeeze operator with squeezing parameter $\nu_{q}=-\frac{1}{4}\ln(1-\zeta^{2}_{q})$ and respectively $\hat{D}(\alpha_{q})$ is the displacement operator with complex amplitude $\alpha_{q}=|\alpha_{q}|e^{i\Phi_{q}}$
\begin{eqnarray}
&&|\alpha_{q}|=\frac{x_{0}F_{q}}{\hbar\omega_{q}2\sqrt{2}}\sqrt{\frac{\cos^{2}(\xi-\phi)}{(1-\zeta_{q}^{2})^{2}}+\sin^{2}(\xi-\phi)},\notag\\
&&\Phi_{q}=\tan^{-1}\{(1-\zeta_{q}^{2})\tan(\xi-\phi)\}.\label{displacement}
\end{eqnarray}
Here the magnitude of the displacement amplitude for the center-of-mass mode $|\alpha_{c}|$ is proportional to the sum of the forces $F_{c}=F_{1}+F_{2}$, while the amplitude $|\alpha_{r}|$ is proportional to the force difference $F_{r}=F_{-}$. The latter implies that by measuring simultaneously the mean phonon number in the center-of-mass mode and in the rocking mode for example by addressing each ion by additional blue-detuned laser field \cite{Haffner2008} one could detect the magnitude of the force as well as its phase. Indeed, let's assume that the system is prepared initially in the vibrational ground state for the both vibrational modes, the expectation value of the phonon number operator for mode $q$ at time $t$ is given by
\begin{eqnarray}
\langle \hat{a}^{\dag}_{q}\hat{a}_{q}\rangle&=&|\alpha_{q}|^{2}\sin(2\Phi_{q})\sinh(2\nu_{q})
(\sin(2\vartheta_{q} t)-2\sin(\vartheta_{q} t))\notag\\
&&-\cos(2\Phi_{q})\sinh(4\nu_{q})\sin^{2}(\vartheta_{q}t)-2|\alpha_{q}|^{2}\cos(\vartheta_{q} t)\notag\\
&& -\frac{1}{2}(1+2|\alpha_{q}|^{2})\cos(2\vartheta_{q}t)\sinh^{2}(2\nu_{q})+c_{q},\label{signal_n}
\end{eqnarray}
where $c_{q}$ is time-independent parameter
\begin{equation}
c_{q}=\frac{1}{2}\{|\alpha_{q}|^{2}\cosh(4\nu_{q})+\sinh^{2}(2\nu_{q})+3|\alpha_{q}|^{2}\}.
\end{equation}
\begin{figure}
\includegraphics[width=0.45\textwidth]{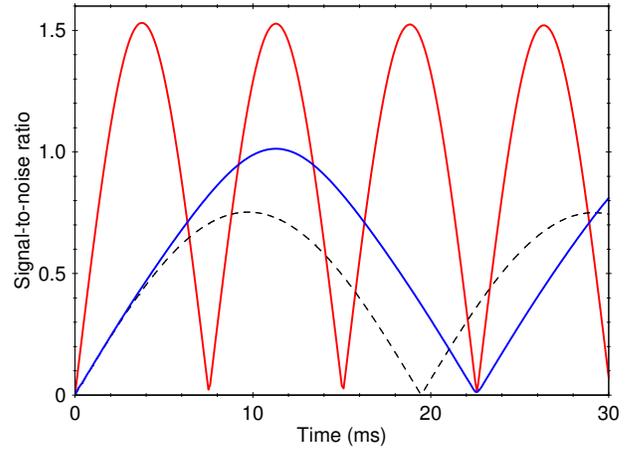}
\caption{(Color online) The exact result of the signal-to-noise ratio versus time. The parameter are set to $F_{1}=7.5$ yN and $F_{2}=5.0$ yN. The red curve is the $\rm SNR$ for the center-of-mass mode and respectively the blue curve is the $\rm SNR$ for the rocking mode. The dashed line shows the $\rm SNR$ for the two coupled harmonic oscillators setting $g=0$. }
\label{var}
\end{figure}

In Figure \ref{fig2} we compared the exact result for the mean-phonon number of the both collective modes with the expression (\ref{signal_n}) as a function of time. Perfect agreement is observed with the analytical and exact results barely discernible. From Eq. (\ref{signal_n}) it is straightforward to show that at time $\vartheta_{q}t_{*}=k_{q}\pi$ with $k_{q}$ being odd number the mean-phonon number is simplified to $\langle \hat{a}^{\dag}_{q}\hat{a}_{q}\rangle=4|\alpha_{q}|^{2}$, where for $\Phi=0$ ($\phi=\xi$) the signal reaches the maximal value. We note that the time $t_{*}$ is different for the center-of-mass mode and rocking mode because $\omega_{c}>\omega_{r}$. The latter implies that in general is not possible the signals for the both vibrational modes to reach their maximal value simultaneously. However, if we set for example $t_{*}=\pi k_{c}/\vartheta_{c}$ one can determine the hopping amplitude $\kappa_{*}$ for which the condition $t_{*}=\pi k_{r}/\vartheta_{r}$ is fulfilled. Indeed, solving the equation $(1-x)(1-x-\zeta^{2})=(k_{r}/k_{c})^{2}(1+x)(1+x-\zeta^{2})$ for $x=\kappa/\delta$ gives the desired hopping amplitude, where $\zeta^{2}=4g^{2}/\Omega\delta$. Note that for $k_{c}>k_{r}$ the equation has always one real positive root $0<\kappa/\delta<1$. In Figure \ref{fig3} we show the the mean phonon number at the time $t_{*}$ as a function of the laser phase $\phi$. By varying the laser phase, the signal oscillate with amplitude proportional to the force difference $F_{-}$ for the rocking mode and respectively to the total force $F_{c}$ for the center-of-mass mode. Moreover, the maximum of the signal correspond to the phase $\phi_{\max}=\xi$ and respectively the minimum to $\phi_{\min}=\xi+\pi/2$ which allows to determine the unknown phase $\xi$ via measuring the mean phonon number versus the laser phase $\phi$.

Next, we discuss the signal-to-noise ratio ${\rm{SNR}}=\langle \hat{a}^{\dag}_{q}\hat{a}_{q}\rangle/\langle \Delta^{2}\hat{a}^{\dag}_{q}\hat{a}_{q}\rangle^{1/2}$ as a figure of merit for the force sensitivity. At time $t_{*}$ we have ${\rm{SNR}}=2|\alpha_{q}|$ which indicates that for higher coupling $\zeta_{q}$ the respective displacement amplitude increases which leads to enhanced force sensitivity, see Fig. \ref{var}. Setting ${\rm{SNR}}$ to one we find the minimal detectable force
\begin{equation}
F_{q}^{\rm min}=\frac{\sqrt{2}\hbar\omega_{q}}{x_{0}}(1-\zeta_{q}^{2}).
\end{equation}
Consider for example the parameters in Fig. \ref{var} the minimal detectable force difference is of order of $2.5$ yN. Note that for a such force difference the corresponding $\rm SNR$ for quantum probe consisting of two-coupled harmonic oscillators with $g=0$ is less than one emphasizing the advantages of the strong spin-boson coupling.

Since we assume that the state vector evolves in time according to unitary propagator $\hat{U}(t)$ it is straightforward to show that quantum Fisher information is given by (see Appendix \ref{fisher_Q} for more details)
\begin{equation}
I_{Q}(\varphi)=16\left(\frac{\partial\alpha_{q}}{\partial\varphi}\right)\left(\frac{\partial\alpha^{*}_{q}}{\partial\varphi}\right),\label{QFI_G}
\end{equation}
where $\varphi$ is either $\varphi=F_{q}$ or $\varphi=\xi$. For the estimation of the spatial variation of the force we set $\xi=\phi$ and $\vartheta_{q}t_{*}=k_{q}\pi$ such that for $\varphi=F_{q}$ we obtain
\begin{equation}
I_{Q}(F_{q})=2\left(\frac{x_{0}}{\hbar\omega_{q}(1-\zeta_{q}^{2})}\right)^{2}.\label{QFI}
\end{equation}
The result (\ref{QFI}) shows that ultimate uncertainty in the force estimation for single experimental realization is given by $\Delta^{2}F_{q}\geq (F_{q}^{\rm min}/2)^{2}$. Moreover, Eq. (\ref{QFI}) indicates that by approaching $\zeta_{q}$ the critical coupling $\zeta_{q}^{\rm c}=1$, the quantum Fisher information diverges such that the system becomes sensitive to infinitely small force perturbation.

We can derive a similar expression for the estimation of the phase of the force. Indeed, from Eq. (\ref{QFI_G}) one can derive the maximal value of the quantum Fisher information for the phase estimation
\begin{equation}
I_{Q}(\xi)=2\left(\frac{x_{0}F_{q}}{\hbar\omega_{q}(1-\zeta_{q}^{2})}\right)^{2},
\end{equation}
which is attained for $\phi=\xi+\pi/2$, (see Appendix  \ref{fisher_Q}).

\begin{table}
\caption{Minimal detectable signal}
\centering
\begin{tabular}{c c c}
\hline\hline
Method & Force & Magnetic Gradient \\ [0.5ex]
\hline

Adiabatic & $F_{-}^{\rm min}=\frac{2\hbar \gamma(\delta-\kappa)}{\pi g x_{0}}\sinh^{-1}(1)$ & $B^{\prime}_{\rm min}=\frac{2\gamma}{\pi\lambda\Delta z}\sinh^{-1}(1)$ \\
C.H.O &$F_{q}^{\rm min}=\frac{\sqrt{2}\hbar\omega_{q}}{x_{0}}(1-\zeta_{q}^{2})$&\\
\hline
\end{tabular}
\label{table}
\end{table}

Finally, we point out that the although the estimation of the number of phonons is experimentally convenient observable, it does not saturate the
quantum Cramer-Rao bound (\ref{QCR}) associated with the quantum Fisher information. The optimal measurements that saturate the Eq. (\ref{QCR})
are projective measurements formed by the eigenvectors of the Symmetric Logarithmic Derivative (SLD) operator $\hat{L}_{\varphi}$ defined by
\begin{equation}
\frac{\partial\hat{\rho}_{\varphi}}{\partial\varphi}=\frac{1}{2}\{\hat{L}_{\varphi}\hat{\rho}_{\varphi}+\hat{\rho}_{\varphi}\hat{L}_{\varphi}\},
\end{equation}
where $\hat{\rho}_{\varphi}$ is the density operator of the system. For pure state the SLD operator can be expressed as
$\hat{L}_{\varphi}=2\{|\psi\rangle\langle\partial_{\varphi}\psi|+|\partial_{\varphi}\psi\rangle\langle\psi|\}$ \cite{Paris2009}.
It is straightforward to show that at time $t_{*}$ we have $\langle\psi|\partial_{\varphi}\psi\rangle=0$ such that the
eigenvectors of $\hat{L}_{\varphi}$ are $|l_{\pm}\rangle=|\partial_{\varphi}\psi\rangle\pm\sqrt{\langle\partial_{\varphi}\psi|\partial_{\varphi}\psi\rangle}|\psi\rangle$
with eigenvalues $l_{\pm}=2\sqrt{\langle\partial_{\varphi}\psi|\partial_{\varphi}\psi\rangle}$ which are independent on the parameter
we wish to estimate.
\section{Conclusions}\label{conclusions}
We have shown that the two coupled harmonic oscillators driven by external laser fields can served as an efficient detector of spatially varying
electric and magnetic fields. We have discussed an adiabatic sensing protocol and show that the small force difference can be detected by
measuring the spin population. We have shown that the information of the phase of the force also is mapped onto the spin states and thus it can
be extracted by the same strategy. The adiabatic sensing technique can be used also for measuring magnetic field gradient. Furthermore, we have shown that the strong spin phonon can be used to improve the force sensitivity. Here the force estimation is performed by measuring the mean-phonon
number of the collective vibrational modes. We have shown that higher spin-phonon
coupling leads to enhance sensitivity to force and phase estimations. We have quantified the estimation uncertainty by using signal-to-noise ratio as a figure of merit for the sensitivity as well as by using the quantum Fisher information. We summarize in Table \ref{table} the corresponding minimal detectable signals for the sensing of spatially varying force and magnetic field gradient using adiabatic method and coupled harmonic system (C. H. S) as a quantum probe. Using realistic experimental parameters we have shown that our sensing protocols can be used to detect forces in the range of few yN as well as magnetic field gradients with magnitude of $\rm{pT}$/$\mu$m.
\acknowledgments

This work has been supported by the DFG through the SFB-TR 185.

\appendix
\section{Solution of the two-state problem}\label{STSP}
The coupled system of differential equations which describe the system in the symmetry-broken phase is given by
\begin{eqnarray}
&&i\frac{d c_{+}}{dt}=-\alpha c_{+}-\Delta_{\rm c}e^{-2\gamma t} c_{-},\notag\\
&&i\frac{d c_{-}}{dt}=\alpha c_{-}-\Delta_{\rm c}e^{-2\gamma t} c_{+},
\end{eqnarray}
which is solved with the initial conditions $c_{+}(0)=c_{-}(0)=\frac{1}{\sqrt{2}}$. Here we set $\alpha=\cos(\phi-\xi)\frac{g x_{0}F_{-}}{\hbar\omega_{r}}$ and $\Delta_{\rm c}=\Omega^{2}/4J$ is the coupling between the states $|\Psi_{+}\rangle$ and $|\Psi_{-}\rangle$ which is obtained by using a second-order degenerate perturbation theory.

The solution of the system can be written as
\begin{eqnarray}
c_{+}(t)&=&\frac{\pi x e^{-\gamma t}}{2\sqrt{2}\cosh\left(\frac{\pi\alpha}{2\gamma}\right)}\{(J_{1-\beta}(x)-i J_{-\beta}(x))J_{\beta}(x e^{-2\gamma t})\notag\\
&&+(J_{\beta-1}(x)+i J_{\beta}(x))J_{-\beta}(x e^{-2\gamma t})\}
\end{eqnarray}
and respectively
\begin{eqnarray}
c_{-}(t)&=&\frac{\pi x e^{-\gamma t}}{2\sqrt{2}\cosh\left(\frac{\pi\alpha}{2\gamma}\right)}\{(J_{-\beta}(x)+i J_{1-\beta}(x))J_{\beta-1}(x e^{-2\gamma t})\notag\\
&&+(J_{\beta}(x)-i J_{\beta-1}(x))J_{1-\beta}(x e^{-2\gamma t})\},
\end{eqnarray}
where $J_{\nu}(y)$ is a Bessel function of the first kind. Here $\beta=\frac{1}{2}+i\frac{\alpha}{2\gamma}$ and $x=\frac{\Delta_{\rm c}}{2\gamma}$. Using the asymptotic expressions $J_{\nu}(y)\sim \frac{(y/2)^{\nu}}{\Gamma(1+z)}$ for $t\gg \gamma^{-1}$ and $J_{\nu}(y)\sim\sqrt{\frac{2}{\pi y}}\cos\left(y-\frac{\nu\pi}{2}-\frac{\pi}{4}\right)$ for $y\gg |v^{2}-1/4|$ we obtain
\begin{eqnarray}
&&c_{+}(t)\approx \sqrt{\frac{\pi}{2}}\frac{e^{i x}}{\cosh\left(\frac{\pi\alpha}{2\gamma}\right)}\frac{z^{-i\frac{\alpha}{2\gamma}}e^{\frac{\pi\alpha}{4\gamma}}}
{\Gamma\left(\frac{1}{2}-i\frac{\alpha}{2\gamma}\right)},\notag\\
&&c_{-}(t)\approx \sqrt{\frac{\pi}{2}}\frac{e^{i x}}{\cosh\left(\frac{\pi\alpha}{2\gamma}\right)}\frac{z^{i\frac{\alpha}{2\gamma}}e^{-\frac{\pi\alpha}{4\gamma}}}
{\Gamma\left(\frac{1}{2}+i\frac{\alpha}{2\gamma}\right)},\label{AS}
\end{eqnarray}
where $z=(x/2)e^{-2\gamma t}$.

The quantum Fisher information can be written as
\begin{eqnarray}
I_{Q}(p)&=&4\left(\frac{gx_{0}}{2\hbar\omega_{r}\gamma}\right)^{2}\{\partial_{p} c^{*}_{+}\partial_{p} c_{+}+\partial_{p} c^{*}_{-}\partial_{p} c_{-}\notag\\
&&-|c^{*}_{+}\partial_{p} c_{+}+c^{*}_{-}\partial_{p} c_{-}|^{2}\},
\end{eqnarray}
where $p=\alpha/2\gamma$. For the force difference estimation using Eq. (\ref{AS}) and setting $\phi=\xi$ we find
\begin{equation}
I_{Q}(F_{-})=\left(\frac{g x_{0}}{2\hbar\gamma\omega_{r}}\right)^{2}\frac{\pi^{2}+4(\ln(z)-\Re \Psi(\beta))^{2}}{\cosh^{2}\left(\frac{\pi gx_{0}F_{-}}{2\hbar\gamma\omega_{r}}\right)}.
\end{equation}
Similar expression can be obtained for the phase estimation
\begin{equation}
I_{Q}(\xi)=\left(\frac{g x_{0}F_{-}\sin(\phi-\xi)}{2\hbar\gamma\omega_{r}}\right)^{2}\frac{\pi^{2}+4(\ln(z)-\Re \Psi(\beta))^{2}}{\cosh^{2}\left(\frac{\pi gx_{0}F_{-}\cos(\phi-\xi)}{2\hbar\gamma\omega_{r}}\right)}.
\end{equation}

\section{Three Ion Case}\label{TIC}
Consider a system of three trapped ions with nearest neighbour hopping amplitude
\begin{equation}
\hat{H}_{x}=\hbar\delta\sum_{j=1}^{3}\hat{a}_{j}^{\dag}\hat{a}_{j}
+\hbar\kappa\sum_{j=1}^{2}(\hat{a}_{j}^{\dag}\hat{a}_{j+1}+{\rm h.c.}).\label{Hopping3}
\end{equation}
With Eq. (\ref{Hopping3}) and for $\Omega=0$ the total Hamiltonian becomes
\begin{equation}
\hat{H}_{\rm RL}=\hbar\delta\sum_{j=1}^{3}\hat{a}_{j}^{\dag}\hat{a}_{j}
+\hbar\kappa\sum_{j=1}^{2}(\hat{a}_{j}^{\dag}\hat{a}_{j+1}+{\rm h.c.})
+\hbar\sum_{j=1}^{3}g_{j}\sigma_{j}^{z}(\hat{a}_{j}^{\dag}+\hat{a}_{j}).
\end{equation}
We introduce collective modes according to the transformation $\hat{a}_{1}=(\hat{a}_{c}+\hat{a}_{r}-\sqrt{2}\hat{a}_{e})/2$, $\hat{a}_{2}=(\hat{a}_{c}-\hat{a}_{r})/\sqrt{2}$ and $\hat{a}_{3}=(\hat{a}_{c}+\hat{a}_{r}+\sqrt{2}\hat{a}_{e})/2$, which diagonalize the hopping Hamiltonian (\ref{Hopping3}) such that $\hat{H}_{x}=\hbar\omega_{c}\hat{a}_{c}^{\dag}\hat{a}_{c}+\hbar\omega_{r}\hat{a}_{r}^{\dag}\hat{a}_{r}+\hbar\omega_{e}\hat{a}_{e}^{\dag}\hat{a}_{e}$ where the collective vibrational frequencies are $\omega_{c}=\delta+\sqrt{2}\kappa$, $\omega_{r}=\delta-\sqrt{2}\kappa$ and $\omega_{e}=\delta$.

Next we assume that $g_{1}=g_{3}=g$, $g_{2}=\sqrt{2}g$ and perform transformation
\begin{equation}
\hat{\tilde{H}}=\prod_{q}\hat{D}^{\dag}(\hat{\alpha}_{q})\hat{H}_{\rm RL}\hat{D}^{\dag}(\hat{\alpha}_{q}),\quad q=c,r,c,
\end{equation}
where $\hat{D}(\hat{\alpha}_{q})$ is a displacement operator with
\begin{eqnarray}
&&\hat{\alpha}_{c}=-\frac{g}{2\omega_{c}}(\sigma_{1}^{z}+2\sigma_{2}^{z}+\sigma_{3}^{z}),\quad \hat{\alpha}_{e}=\frac{g}{\sqrt{2}\omega_{e}}(\sigma_{1}^{z}-\sigma_{3}^{z}),\notag\\
&&\hat{\alpha}_{r}=-\frac{g}{2\omega_{r}}(\sigma_{1}^{z}-2\sigma_{2}^{z}+\sigma_{3}^{z}).
\end{eqnarray}
Using this we find
\begin{eqnarray}
\hat{\tilde{H}}&=&\hbar\omega_{c}\hat{a}_{c}\hat{a}_{c}+\hbar\omega_{r}\hat{a}_{r}\hat{a}_{r}+\hbar\omega_{e}\hat{a}^{\dag}_{e}\hat{a}_{e}
+\hbar J(\sigma_{1}^{z}\sigma_{2}^{z}+\sigma_{2}^{z}\sigma_{3}^{z})\notag\\
&&-\hbar J^{\prime}\sigma^{z}_{1}\sigma^{z}_{3}.
\end{eqnarray}
The Hamiltonian describes interaction between three spins with coupling strengths
\begin{equation}
J=g^{2}\left(\frac{1}{\omega_{r}}-\frac{1}{\omega_{c}}\right),\quad J^{\prime}=g^{2}\left(\frac{1}{2\omega_{r}}+\frac{1}{2\omega_{c}}-\frac{1}{\omega_{e}}\right).
\end{equation}
Since $J,J^{\prime}>0$ the ground state spin order in the original basis is given by
\begin{equation}
|\Psi_{+}\rangle=\left|\downarrow\uparrow\downarrow\right\rangle|0_{c}\rangle|\alpha_{r}\rangle|0_{e}\rangle,\quad |\Psi_{-}\rangle=\left|\uparrow\downarrow\uparrow\right\rangle|0_{c}\rangle|-\alpha_{r}\rangle|0_{e}\rangle,
\end{equation}
where $|\alpha_{r}\rangle$ is a coherent state in the rocking mode with displacement amplitude $\alpha_{r}=2g/\omega_{r}$.
\section{Quantum Fisher Information}\label{fisher_Q}

Consider that the system is prepared initially in the state $|\psi(0)\rangle=\left|\downarrow\downarrow\right\rangle|0_{c}\rangle|0_{r}\rangle$ and evolves in time according to $|\psi(t)\rangle=\hat{U}_{c}(t)\hat{U}_{r}(t)|\psi(0)\rangle$, where $\hat{U}_{q}(t)=\hat{D}^{\dag}(\alpha_{q})\hat{S}^{\dag}(\nu_{q})e^{-i\vartheta_{q}t\hat{a}_{q}^{\dag}\hat{a}_{q}}\hat{S}(\nu_{q})\hat{D}(\alpha_{q})$. Using the properties
\begin{eqnarray}
&&\hat{D}(\alpha_{q})\hat{a}_{q}\hat{D}^{\dag}(\alpha_{q})=\hat{a}_{q}-\alpha_{q},\notag\\
&&\hat{S}(\nu_{q})\hat{a}_{q}\hat{S}^{\dag}(\nu_{q})=\hat{a}_{q}\cosh(\nu_{q})+\hat{a}_{q}^{\dag}\sinh(\nu_{q}),\label{properties}
\end{eqnarray}
we find
\begin{equation}
\langle \psi|\partial_{F_{q}}\psi\rangle=-2i\alpha_{q}(\partial_{F_{q}}\alpha_{q})e^{-2\nu_{q}}\sin(\vartheta_{q}t),\label{1}
\end{equation}
where we assume that $\alpha_{q}$ is real. Next we write
\begin{eqnarray}
\langle\partial_{F_{q}}\psi|\partial_{F_{q}}\psi\rangle&=&(\partial_{F_{q}}\alpha_{q})^{2}\{2(1-\cos(\vartheta_{q}t))+
2\langle \hat{a}^{\dag}_{q}\hat{a}_{q}\rangle\notag\\
&&-2\alpha_{q}^{2}(1-2\cos(\vartheta_{q}t))-A_{q}-A_{q}^{*}\},\label{2}
\end{eqnarray}
where
\begin{eqnarray}
A_{q}&=&\langle 0_{q}|\hat{D}^{\dag}(\alpha_{q})\hat{S}^{\dag}(\nu_{q})e^{i\vartheta_{q}t\hat{a}_{q}^{\dag}\hat{a}_{q}}
\hat{S}(\nu_{q})\hat{a}_{q}^{2\dag}\hat{S}^{\dag}(\nu_{q})e^{-i\vartheta_{q}t\hat{a}_{q}^{\dag}\hat{a}_{q}}\notag\\
&&\times\hat{S}(\nu_{q})\hat{D}(\alpha_{q})|0_{q}\rangle.
\end{eqnarray}
Using Eq. (\ref{properties}) we obtain
\begin{eqnarray}
A_{q}&=&i\sin(\vartheta_{q}t)\sinh(2\nu_{q})(e^{-i\vartheta_{q}t}\sinh^{2}(\nu_{q})-e^{i\vartheta_{q}t}\cosh^{2}(\nu_{q}))\notag\\
&&+\alpha_{q}^{2}e^{-2\nu_{q}}(e^{i\vartheta_{q}t}\cosh(\nu_{q})+e^{-i\vartheta_{q}t}\sinh(\nu_{q}))^{2}.
\end{eqnarray}
The quantum Fisher information oscillates with time and reach maximal value at $\vartheta_{q}t_{*}=k_{q}\pi$ with $k_{q}$ odd number. Using Eqs. (\ref{1}) and (\ref{2}) we find
\begin{equation}
I_{Q}(F_{q})=16\left(\frac{\partial\alpha_{q}}{\partial F_{q}}\right)^{2},
\end{equation}
which is exactly the result (\ref{QFI}). The expression can be generalized for complex amplitude $\alpha_{q}$. Following the same steps as above the quantum Fisher information at $\vartheta_{q}t_{*}=k_{q}\pi$ is given by
\begin{equation}
I_{Q}(\varphi)=16\left(\frac{\partial\alpha_{q}}{\partial\varphi}\right)\left(\frac{\partial\alpha^{*}_{q}}{\partial\varphi}\right),
\end{equation}
where $\varphi$ is either $\varphi=F_{q}$ or $\varphi=\xi$. Finally, setting $\varphi=\xi$ and using Eq. (\ref{displacement}) we find
\begin{equation}
I_{Q}(\xi)=2\left(\frac{F_{q}x_{0}}{\hbar\omega_{q}}\right)^{2}\left(\cos^{2}(\xi-\phi)+\frac{\sin^{2}(\xi-\phi)}{(1-\zeta_{q}^{2})^{2}}\right),
\end{equation}
which reaches the maximal value at $\phi=\xi+\pi/2$.

\end{document}